\documentclass[aps,pra,10pt,showpacs,amsmath,amssymb,floatfix,superscriptaddress,longbibliography,twocolumn,eqsecnum]{revtex4-2}
\usepackage{physics}
\usepackage{graphicx}
\usepackage[usenames,dvipsnames,svgnames,table]{xcolor}
\usepackage{tcolorbox}
\usepackage[unicode=true,
            pdfusetitle,
            bookmarks=true,
            bookmarksnumbered=false,
            bookmarksopen=false,
            breaklinks=true,
            pdfborder={0 0 0},
            backref=false,
            colorlinks=true,
            hypertexnames=false]{hyperref}
\hypersetup{linkcolor=NavyBlue,urlcolor=NavyBlue,citecolor=NavyBlue}

\newcommand{\qfi}{\mathcal{F}}
\newcommand{\revise}[1]{#1}

\begin{document}
\title{Performance-tradeoff relation for locating two incoherent optical point sources}
\author{Jingjing Shao}
\affiliation{School of Sciences, Hangzhou Dianzi University, Hangzhou 310018, China}
\author{Xiao-Ming Lu}
\email{lxm@hdu.edu.cn}
\homepage{http://xmlu.me}
\affiliation{School of Sciences, Hangzhou Dianzi University, Hangzhou 310018, China}

\begin{abstract}
	The optimal quantum measurements for estimating individual parameters might be incompatible with each other so that they cannot be jointly performed.
	The tradeoff between the estimation precision for different parameters can be characterized by information regret---the difference between the Fisher information and its quantum limit.
	We show that the information-regret-tradeoff relation can give us not only an intuitive picture of the potential in improving the joint scheme of estimating the centroid and the separation, but also some clues to the optimal measurements for the sequential scheme.
	In particular, we show that, for two incoherent point sources with a very small separation, the optimal measurement for the separation must extract little information about the centroid, and vice versa.
\end{abstract}

\maketitle

\section{Introduction} 
\label{sec:introduction}
The resolution power of incoherent imaging plays an important role in astronomy observation and biological imaging.
The conventional criterion on resolving two incoherent point sources---the Rayleigh criterion~\cite{LordRayleigh1879}---is not a rigorous approach and cannot be used to deal with the randomness from quantum measurements~\cite[Sec.~30-4]{Feynman1963}.
A modern method of assessing the resolution power utilizes statistical approaches like parameter estimation or hypothesis testing~\cite{Harris1964,Ram2006,Chao2016,Dekker1997,Tsang2016b,Sidhu2017,Tsang2018}.
Combining these statistical approaches with quantum mechanics, Tsang \textit{et~al}.~\cite{Tsang2016b} recently revealed that the resolution power of incoherent imaging in principle can be significantly improved by optimizing the quantum measurement performed on the far field.
Concretely, spatial mode demultiplexing (SPADE) has a much better performance than direct imaging for resolving two incoherent point sources~\cite{Tsang2016b}. 
Since then, many efforts have been devoted to pursuing the super-resolution of incoherent imaging, e.g., experimental implementations~\cite{Tham2017,Paur2016,Yang2016,Tang2016,Parniak2018,Paur2018,Zhou2019a,Paur2019,Boucher2020,Wadood2021,Larson2021,Santra2021,Pushkina2021} and some theoretic generalizations~\cite{Nair2016,Nair2016a,Tsang2017a,Lupo2016,Rehacek2017a,Lu2018,Tsang2019,Zhou2019,Peng2021,Wang2021,Len2020,Datta2020} (see Ref.~\cite{Tsang2019a} for a recent review).

Remarkably, the SPADE measurement, which is designed for estimating the separation between two point sources, requires prior information about the centroid of two point sources and the alignment of the device with the centroid~\cite{Tsang2016b}.
The prior information about the centroid can be readily obtained, as direct imaging is accurate in estimating the centroid.
Grace \textit{et~al}. proposed an adaptive two-stage detection scheme that dynamically allocates the resources between the centroid estimation and the separation estimation and showed that it outperforms direct imaging~\cite{Grace2020}.
de Almeida \textit{et~al}. investigated the impact of the misalignment of the SPADE device on separation estimation~\cite{Almeida2021}.
These works on the sequential estimation for the centroid and the separation for two incoherent point sources demonstrate the feasibility of SPADE.

The centroid and separation of two incoherent point sources can also be jointly estimated by repeating the same quantum measurement on samples.
The performance of a quantum measurement for the joint estimation of multiple parameters can be assessed by the Fisher information matrix.
However, the present-day theoretic tools for studying the quantum limit of multiparameter estimation---mainly, the quantum Cram\'er-Rao bounds~\cite{Helstrom1967,Helstrom1968,Helstrom1976,Yuen1973,Holevo2011,Lu2020a,Liu2020,Polino2020}---suffer from the incompatibility issue of quantum measurements~\cite{Lu2021,Ragy2016,Carollo2019a,Suzuki2016,Kull2020,Sidhu2020,Candeloro2021}.
Chrostowski et~al.~\cite{Chrostowski2017} showed that the optimal measurements for individually estimating the centroid and the separation of two incoherent point sources are incompatible.
Parniak \textit{et~al}.~\cite{Parniak2018} implemented a joint estimation of the centroid and the separation of two incoherent point sources near the quantum limit by using collecting measurements on two photons.
It remains open what tradeoff relation between the estimation precision of the centroid and that of the separation should be obeyed when taking optimization over quantum measurements.

Recently, one of the present authors and Wang~\cite{Lu2021} introduced the concept of information regret to characterize the performance of a quantum measurement for estimating individual parameters and established a tradeoff relation between the information regrets for any two parameters of interest. 
This information regret tradeoff relation (IRTR) supplies us with a useful theoretic tool for investigating the multiparameter estimation problem of locating two incoherent optical point sources.
In this work, we will show that the IRTR can give us an intuitive picture of the possible improvements of quantum measurements for not only the joint scheme but also the sequential scheme of estimating the centroid and the separation.
Moreover, for the case of maximum incompatibility, the IRTR gives an important clue to the optimal quantum measurement for one individual parameter of interest, provided that plenty of prior information about other parameters is available.
Particularly, we will show that, for two incoherent point sources with a very small separation, the optimal measurement for the separation must extract little information about the centroid, and vice versa.

This paper is organized as follows.
In Sec.~\ref{sec:model}, we give a brief introduction on the model of locating two incoherent optical point sources, the method of quantum parameter estimation, and the information regret tradeoff relation.
In Sec.~\ref{sec:IRTR_analysis}, we study the IRTR for quantum estimation problem underlying the resolution of two incoherent optical point sources.
In Sec.~\ref{sec:measurement}, we use the IRTR to assess several quantum measurements as well as randomly generated ones for estimating the centroid and the separation of two sources.
We summarize our work in Sec.~\ref{sec:conclusion} and give the details of some calculations in the Appendix~\ref{sec:calculation_incomp}.

\section{Parameter estimation for two incoherent point sources} 
\label{sec:model}
\subsection{Model}

Focusing on the spatial resolution, we follow Ref.~\cite{Tsang2016b} to consider the imaging system with quasimonochromatic scalar paraxial waves, one spatial dimension, and two equally-bright incoherent point sources.
Assuming that the sources are weak, the one-photon state of the optical field on the image plane provides the dominant information for locating the positions of the sources.
The density operator describing the one-photon state with a diffraction-limited imaging system can be written as~\cite{Tsang2016b} 
\begin{align}
	\rho & =\frac12 \qty(\op{\psi_1}+\op{\psi_2}), \label{eq:one_photon}\\
	\ket{\psi_j} &= \int \dd x\, \psi_j(x) \ket x \text{ for } j=1,2, 
\end{align}
where \(x\) is the image-plane coordinate, \(\ket x\) denotes the image-plane position eigenket for a photon, and \(\psi_j(x)\) represents the image-plane wave function of a photon from the \(j\)-th source.
Furthermore, assume that the imaging system is spatially invariant such that 
\begin{equation}
	\psi_j(x)=\psi(x - X_j),	
\end{equation}
where \(\psi(x)\) is the normalized point-spread function of the imaging system and \(X_j\) is the unknown position of each source. 
The resolution of the two optical point sources can treated as the problem of estimating the parameters \(X_1\) and \(X_2\).

\subsection{Parameter estimation theory}

We now briefly review the quantum multiparameter estimation theory.
Suppose that the density operator \(\rho\) depends on an unknown vector parameter \(\theta=(\theta_1,\theta_2,\ldots,\theta_n)\).
Denote by \(\hat\theta_j\) an estimator for \(\theta_j\), which represents the data-processing and maps the observation data to the estimates.
The estimation error can be characterized by the error-covariance matrix \(\mathcal E\) defined by 
\begin{equation}
	\mathcal E_{jk} := \mathbb{E}_{\theta}\qty[
		(\hat{\theta}_{j} - \theta_{j})
		(\hat{\theta}_{k} - \theta_{k})
	],
\end{equation}
where the expectation \(\mathbb{E_{\theta}\left[\bullet\right]}\) is taken with respect to the probability distribution of observation data.
A quantum measurement is described by a positive-operator-valued measure (POVM): 
\(M = \qty{M_w \mid M_w \geq 0, \sum_w M_w = \openone}\) with \(w\) denoting the outcome and \(\openone\) the identity operator.  
The probability of obtaining an outcome \(\omega\) is given by \(p(w) = \tr(M_w\rho)\) according to Born's rule in quantum mechanics.
For all unbiased estimators, the error-covariance matrix obeys the Cram\'er-Rao bound (CRB)~\cite{Cramer1946,Rao1945}:
\begin{equation} \label{eq:crb}
	\mathcal{E} \geq \nu^{-1} F(M)^{-1}, 
\end{equation}
where \(\nu\) is the number of experimental repetitions and \(F(M)\) is the Fisher information matrix (FIM)~\cite{Fisher1922} under a quantum measurement \(M\).
The entries of the FIM are defined by
\begin{equation} \label{eq:cfim}
 	F(M)_{jk} := \sum_w \frac1{p(w)} \pdv{p(w)}{\theta_j} \pdv{p(w)}{\theta_k}.
\end{equation}
Note that Eq.~\eqref{eq:crb} is a matrix inequality and should be interpreted in the sense of the Loewer order, i.e., for two matrices \(A\) and \(B\), we say that \(A \geq B\) if \(A - B\) is positive semi-definite. 
The CRB can be asymptotically attained by the maximum likelihood estimator, whose probability tends to be normal with the covariance matrix being \(\nu^{-1}F(M)^{-1}\). 
Therefore, the FIM characterize the performance of a quantum measurement for estimating multiple parameters.

Quantum parameter estimation aims at the optimization of estimation precision over quantum measurements.
For any quantum measurement, the FIM is bounded as~\cite{Braunstein1994,Hiai2014}: 
\begin{equation} \label{eq:bc}
	F(M) \leq \qfi,
\end{equation}
where \(\qfi\) is the quantum Fisher information matrix (QFIM)~\cite{Helstrom1967,Helstrom1968,Helstrom1976,Liu2020,Liu2014a} defined by
\begin{equation}
	\qfi_{jk} := \Re\tr(L_j L_k \rho)
\end{equation}
and \(L_j\), called the symmetric logarithmic derivative (SLD) operator about \(\theta_j\), is the bounded Hermitian operator satisfying 
\begin{equation}
	\pdv{\rho}{\theta_j} = \frac12 \qty(L_j \rho + \rho L_j).
\end{equation}
However, the elements of the FIM in general cannot be maximized simultaneously, known as the incompatibility problem of quantum multiparameter estimation~\cite{Lu2021,Crowley2014,Vidrighin2014,Carollo2019a,Szczykulska2016,Demkowicz-Dobrzanski2020,Sidhu2021,Ragy2016,Suzuki2019,Kull2020}.

\subsection{Regret of Fisher information}



The efficiency of a quantum measurement for multiparameter estimation can be characterized by the information regret matrix introduced in Ref.~\cite{Lu2020}, that is,
\begin{equation}
	R(M) = \qfi - F(M).
\end{equation}
The information regret matrix is always positive semi-definite according to Eq.~\eqref{eq:bc}.
We henceforth call the \(j\)-th diagonal element of \(R(M)\) the \emph{information regret} about the parameter \(\theta_j\).
It will be convenient to use the normalized-square-root (NSR) information regret, which is defined by
\begin{equation} \label{eq:Delta}
	\Delta_{j}
	:=\sqrt{\frac{R_{jj}}{\qfi_{jj}}}
	=\sqrt{\frac{\qfi_{jj}-F_{jj}}{\qfi_{jj}}},
\end{equation}
to express the tradeoff relation between the information regrets regarding different parameters.
Note that the NSR information regret $\Delta_{j}$ must take values in the interval \([0,1]\).

The information regrets about different parameters in general cannot simultaneously be diminished to zero by optimizing over quantum measurements;
They obeys an information regret tradeoff relation (IRTR)~\cite{Lu2021}:
\begin{equation} \label{eq:regret_tradeoff}
	\Delta_1^2 + \Delta_2^2 + 2 \sqrt{1-\tilde{c}^2} \Delta_1 \Delta_2 \geq \tilde{c}^2,
\end{equation}
where \(\tilde c\) is a real coefficient defined as
\begin{equation}  \label{eq:ctilde}
	\tilde{c} = \frac{\tr|\sqrt\rho [L_1, L_2] \sqrt\rho|}
	{2 \sqrt{\qfi_{11} \qfi_{22}}}
\end{equation}
with $\abs{X} := \sqrt{X^{\dagger}X}$ for an operator $X$.
This IRTR was derived by first establishing a correspondence relationship between the information regrets and the state-dependent measurement error defined by Ozawa and then invoking the inequalities that Branciard~\cite{Branciard2013} and Ozawa~\cite{Ozawa2014} proved for formulating the measurement uncertainty relations.
For pure states, the bound Eq.~\eqref{eq:regret_tradeoff} can be saturated by a quantum measurement due to a property of Branciard's inequality~\cite{Branciard2013}.

The IRTR Eq.~\eqref{eq:regret_tradeoff} still holds if replacing \(\tilde c\) by 
\begin{equation} \label{eq:c}
	c := \frac{\abs{\tr([L_1,L_2]\rho)}}{2 \sqrt{\qfi_{11} \qfi_{22}}}. 
\end{equation}
The resultant relation is weaker than Eq.~\eqref{eq:regret_tradeoff}, as we always have \(\tilde c \geq c\) for the same parametric density operator. 
The difference between the two versions of IRTRs can be understood by considering the situations where collective measurements are permitted on \(n\) independent and identically distributed quantum systems.
Denote by \(\tilde c_n\) the incompatibility coefficient calculated for the density operator \(\rho^{\otimes n}\) of \(n\) copies of quantum states.
It has been shown in Ref.~\cite{Chen2021} that 
\begin{equation}
	\lim_{n\to \infty} \tilde c_n = c.
\end{equation}
By the concept of hierarchical incompatibility measures discussed in Ref.~\cite{Chen2021}, we can say that \(\tilde c_1\)---will be simply denoted by \(\tilde c\)---reflects the incompatibility when only independent measurements on each copies are permitted, while \(c\) reflects the incompatibility when all collective measurements on multiple copies are permitted.

\revise{
The incompatibility coefficients \(c\) and \(\tilde c\) can be connected to another incompatibility measure proposed in Ref.~\cite{Carollo2019a}, namely, 
\begin{equation}
	\gamma := \norm{2i\qfi^{-1} \mathcal U}_\infty,
\end{equation}
where \(\mathcal U_{jk} := - (i/4) \tr(\rho [L_j, L_k])\) and \(\norm{\bullet}_\infty\) represents the largest eigenvalue of a matrix.
Assuming that the QFIM is diagonal, for two parameter estimation problems, it was shown that~\cite{Carollo2019a}  
\begin{equation}
	\gamma = \sqrt{\frac{\det 2\mathcal U}{\det\qfi} }.
\end{equation}
Comparing with Eqs.~\eqref{eq:c} and \eqref{eq:ctilde}, we have
\begin{equation}
	\gamma = c \leq \tilde c.
\end{equation}
}

\revise{
\subsection{Estimation error tradeoff}
In terms of the estimation errors, the IRTR implies that~\cite{Lu2021}
\begin{equation}
	\gamma_j +\gamma_k - 2 \sqrt{1-\tilde c_{jk}^2} \sqrt{(1-\gamma_j)(1-\gamma_k)}
	\leq 2 - \tilde c_{jk}^2,
\end{equation}
where \(\gamma_j := 1 / (\nu \mathcal E_{jj} \qfi_{jj})\) with \(\nu\) being the number of experiment repetitions.
Note that \(\gamma_j\) is proportional to the inverse of the estimation error \(\mathcal E_{jj}\) and rescaled to the range \([0,1]\) due to the inequality \(\mathcal E_{jj} \geq 1 / (\nu \qfi_{jj})\).
Particularly, we have
\begin{equation} \label{eq:unit_tilde_c}
	\frac{1}{\nu \mathcal E_{jj} \qfi_{jj} } + 
	\frac{1}{\nu \mathcal E_{kk} \qfi_{kk} }
	\leq 1,
\end{equation}
when \(\tilde c_{jk} = 1\).
This inequality will be used later in our analysis on the joint estimation of the centroid and the separation of two incoherent point optical sources.
}
\section{IRTR analysis for two incoherent point sources} 
\label{sec:IRTR_analysis}

Before using the IRTR to analyze the resolution of two incoherent optical point sources, let us give a brief introduction on the QFIM obtained in Ref.~\cite{Tsang2016b}. 
For the problem of locating two equally-bright optical point sources, it is convenient to use the centroid and the separation as the parameters of interest, that is,
\begin{equation}
	\theta_1 = \frac{X_1 + X_2}{2}
	\qand
	\theta_2 = X_2 - X_1.
\end{equation}
Assume hereafter that the point-spread function \(\psi(x)\) is real.
It has been shown in Ref.~\cite{Tsang2016b} that the QFIM for \(\theta_1\) and \(\theta_2\) is given by
\begin{align}
	\qfi = \mqty(
		4 \kappa - 4 \gamma^2 & 0 \\
		0 & \kappa
	),
\end{align}
where \(\kappa\) and \(\gamma\) are real numbers defined by
\begin{align}
	\kappa  & := \int_{-\infty}^\infty \dd x\, \qty[ \pdv{\psi(x)}{x}]^2, \label{eq:kappa}\\
	\gamma 	& := \int_{-\infty}^\infty \dd x\, \pdv{\psi(x)}{x} \psi(x - \theta_2). \label{eq:gamma}
\end{align}

\subsection{Incompatibility coefficient}
Now, we use the IRTR to analyze the estimation of the centroid and the separation of two weak incoherent optical point sources.
For real point-spread functions, it is easy to see that \(c=0\) (see Appendix~\ref{sec:calculation_incomp} for the details).
Therefore, if collective measurements on multiple samples are permitted, the regrets of Fisher information for both the centroid and the separation can simultaneously vanish.
After some algebras, we get the incompatibility coefficient without considering the collective measurements (see Appendix~\ref{sec:calculation_incomp} for the details):
\begin{equation} \label{eq:incompatibility}
	\tilde c^2 = \frac{\beta^2}{\kappa(\kappa - \gamma^2)},
\end{equation}
where \(\kappa\) and \(\gamma\) are given by Eq.~\eqref{eq:kappa} and Eq.~\eqref{eq:gamma}, respectively, and \(\beta\) is given by
\begin{align} \label{eq:beta}
	\beta & := \int_{-\infty}^\infty \dd x\, \pdv{\psi(x - X_1)}{X_1} \pdv{\psi(x- X_2)}{X_2} \nonumber\\
	&= \int_{-\infty}^\infty \dd x\, \pdv{\psi(x)}{x} \pdv{\psi(x - \theta_2)}{x}.
\end{align}

It is easy to see that the incompatibility coefficient \(\tilde c\) vanishes when \(\beta=0\) is satisfied.
From the definition of \(\beta\), we can see that \(\beta=0\) is equivalent to the fact that the vectors \(\partial\ket{\psi_1} / \partial X_1\) and \(\partial \ket{\psi_2} / \partial X_2\) are orthogonal.
It is known that \(\tilde c = 0\) is necessary for the saturation of Eq.~\eqref{eq:bc} and thus for the vanishing of the regret matrix~\cite{Yang2019b,Chen2021}, however, it remains open whether \(\tilde c = 0\) is also sufficient for the saturation of Eq.~\eqref{eq:bc}.

The most important situation is the estimation of small separations, for which direct imaging has a poor performance and optimal measurement is in demand.
Let us consider the limit of \(\theta_2 \to 0\), i.e., the separation between the two point sources is infinitesimal.
It can be shown from Eq.~\eqref{eq:incompatibility} that 
\begin{equation}
	\lim_{\theta_2 \to 0} \tilde c = 1,
\end{equation}
as \(\lim_{\theta_2 \to 0} \beta = \kappa\) and \(\lim_{\theta_2 \to 0} \gamma = 0\) according to their definitions given in Eq.~\eqref{eq:kappa}, Eq.~\eqref{eq:gamma}, and Eq.~\eqref{eq:beta}.
Therefore, the incompatibility coefficient approaches to its maximum value as the separation decreases to zero.
For \(\tilde c=1\), the IRTR becomes
\begin{equation} \label{eq:maximum_IRTR}
	\Delta_1^2 + \Delta_2^2 \geq 1,
\end{equation}
implying that zero information regret about one parameter must lead to the maximal information regret about the other one.
\revise{We can also use Eq.~\eqref{eq:unit_tilde_c} to get the tradeoff relation about the estimation errors.
As a result, 
\begin{equation}
	\frac{1}{4 \nu \kappa \mathcal E_{11} } + 
	\frac{1}{\nu \kappa \mathcal E_{22} }
	\leq 1
\end{equation}
must hold in the limiting case of \(\theta_2 \to 0\).}
When the incompatibility coefficient attains its maximum value, the resultant IRTR Eq.~\eqref{eq:maximum_IRTR} gives us an important clue to optimize quantum measurements for a parameter of interest.
In such a case, a quantum measurement that attains \(\Delta_2=0\) must satisfy \(\Delta_1=1\), which means that the classical Fisher information about \(\theta_1\) must vanish.
Since \(F_{11} = \sum_\omega \qty[\partial p(\omega) / \partial \theta_1]^2 / p(\omega)\), \(F_{11}\) vanishes if and only if
\begin{equation} \label{eq:necessary}
	\pdv{p(\omega)}{\theta_1} = 0\quad \forall\, \omega \mbox{ such that } p(\omega) \neq 0,
\end{equation}
which in turn is equivalent to
\begin{equation}
	\tr(M_\omega \pdv{\rho}{\theta_1}) = 0 
	\quad \forall\, \omega \mbox{ such that } \tr(M_\omega \rho) \neq 0.
\end{equation}
This condition is necessary for attaining the maximum Fisher information about the separation \(\theta_2\) when the separation between two coherent sources are very small.
\revise{
	Concretely, in the limiting case of \(\theta_2\to0\), if we take \(\qty{\ket \omega}\) as the measurement basis, i.e., \(M_\omega = \op{\omega}\), the above condition is reduced to
	\begin{equation}
		\Re \ip{\omega}{\Psi'_{\theta_1}} \ip{\Psi_{\theta_1}}{\omega} = 0, 	
	\end{equation} 
	where \(\ket{\Psi_{\theta_1}} = \int \psi(x - \theta_1) \ket x \dd x\) and \(\ket{\Psi'_{\theta_1}} = \int \psi'(x - \theta_1) \ket x \dd x\) with \(\psi'(x)\) being the first order derivative of the point-spread function \(\psi(x)\).
	In this way, it seems natural to include the fundamental mode \(\ket{\Psi_{\theta_1}}\) and the first derivative mode \(\ket{\Psi'_{\theta_1}}\) in the optimal measurement basis for estimating very small separations; 
	Such a scheme was already used in Ref.~\cite{Paur2016}.
}
On the other hand, when the separation is very small, the optimal measurement for estimating the centroid must extract little information about the separation.
This is consistent with performance of direct imaging.

\subsection{Gaussian point-spread function}
\begin{figure}[tb]
	\centering
	\includegraphics[]{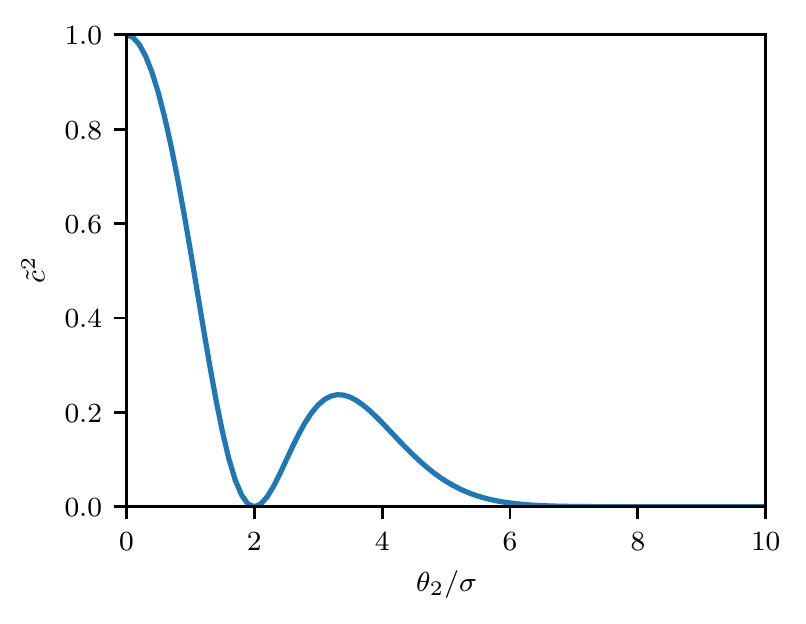}
	\caption{The incompatibility coefficient versus the separation for a Gaussian point-spread function.}
	\label{fig:incomp}
\end{figure}

We now take the Gaussian point-spread function as a typical example, for which we set
\begin{equation} \label{eq:psf_gaussian}
	\psi(x) = \qty(2\pi\sigma^{2})^{-1/4} \exp(- \frac{x^2}{4\sigma^2})
\end{equation}
with \(\sigma\) being a characteristic length of the point-spread function.
It then follows from Eq.~\eqref{eq:incompatibility} that (see Appendix~\ref{sec:calculation_incomp} for the details)
\begin{equation} \label{eq:incompatibility_gaussian}
	\tilde c^2 = 
	\qty(1 - \dfrac{\theta_2^2}{4\sigma^2})^2
	\qty[\exp(\dfrac{\theta_2^2}{4 \sigma^2}) - \dfrac{\theta_2^2}{4\sigma^2}]^{-1}. 
\end{equation}
Figure~\ref{fig:incomp} plots the incompatibility coefficient as a function of \(\theta_2 / \sigma\).
It can be seen that \(\tilde c\) progressively attains its maximum value \(1\) as \(\theta_2\to0\).

Interestingly, there are two cases where the incompatibility coefficient \(\tilde c\) vanishes.
In the first case, at large enough separations, e.g., \(\theta_2 > 6\,\sigma\) in Fig.~\ref{fig:incomp}, \(\tilde c\) approximately vanishes.  
This can be understood by noticing that the one-photon wave functions corresponding to the two point sources are far away from each other so that the locations of the two point sources can be estimated independently from different regions in the image plane.
Consequently, the centroid and the separation can be jointly estimated with the optimal precision. 
The second case of zero incompatibility coefficient is at the specific value \(\theta_2 = 2 \sigma\).
It is still open whether we can find an optimal measurement attaining the maximum Fisher information about \(\theta_1\) and \(\theta_2\) simultaneously when the incompatibility coefficient vanishes.
\revise{
Unlike the case of large enough separation, when \(\theta_2 = 2 \sigma\), the derivatives of the state vectors, \(\pdv*{\ket{\psi_1}}{X_1}\) and \(\pdv*{\ket{\psi_2}}{X_2}\) are orthogonal but \(\ket{\psi_1}\) and \(\ket{\psi_2}\) are not orthogonal;
Thus, the measurement and estimation of \(X_1\) and \(X_2\) will interfere with each other.
It is nontrivial if there exists a measurement that is simultaneously optimal for estimating both the centroid and the separation in such a case.
The vanishing incompatibility coefficient here makes it promising.
}

\section{Measurements for joint estimation} 
\label{sec:measurement}
Now, we study the performance of some specific quantum measurements for estimating the centroid and the separation of two incoherent point sources with the help of IRTR.

\subsection{Direct imaging}
\begin{figure}[bt]
	\centering
	\includegraphics[]{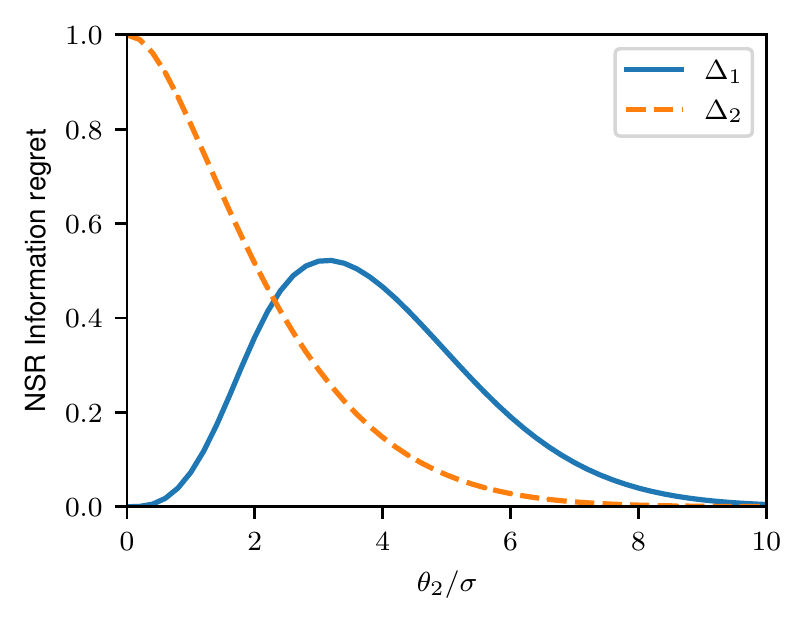}
	\caption{The NSR Information regret versus the separation for direct imaging.}
	\label{fig:regret_direct_imaging}
\end{figure}

Let us first consider direct imaging, which measures the intensity distribution of the optical field on the image plane.
The probability density function of observing a photon at the position \(x=\omega\) is given by
\begin{equation}
	p(\omega) = \frac12 \qty[|\psi(\omega-X_1)|^2 + |\psi(\omega-X_2)|^2],
\end{equation}
where \(X_1\) and \(X_2\) are related to the centroid \(\theta_1\) and the separation \(\theta_2\) as \(X_1 = \theta_1 - \theta_2 / 2\) and \(X_2 = \theta_1 + \theta_2 / 2\).
For Gaussian point-spread functions, we numerically compute the FIM  about the parameters \(\theta_1\) and \(\theta_2\) via Eq.~\eqref{eq:cfim} and then their NSR information regrets via Eq.~\eqref{eq:Delta}.
Figure~\ref{fig:regret_direct_imaging} plots the NSR information regrets of direct imaging for estimating the centroid and the separation of two incoherent optical point sources.
It can be seen that the information regret for estimating the separation of the two sources are large in the sub-Rayleigh region, which was known in the previous work~\cite{Ram2006} and dubbed as Rayleigh's curse~\cite{Tsang2016b,Peng2021}.
Note that for each parameter, its information regret can be reduced to zero.
So Fig.~\ref{fig:regret_direct_imaging} in fact reflects the possible improvements implied by the quantum CRB. 
However, reducing the information regret for one parameter may be accompanied by the inevitable increase of the information regret for the other parameter, with which we need the IRTR to deal. 

\begin{figure*}[tb]
	\includegraphics[]{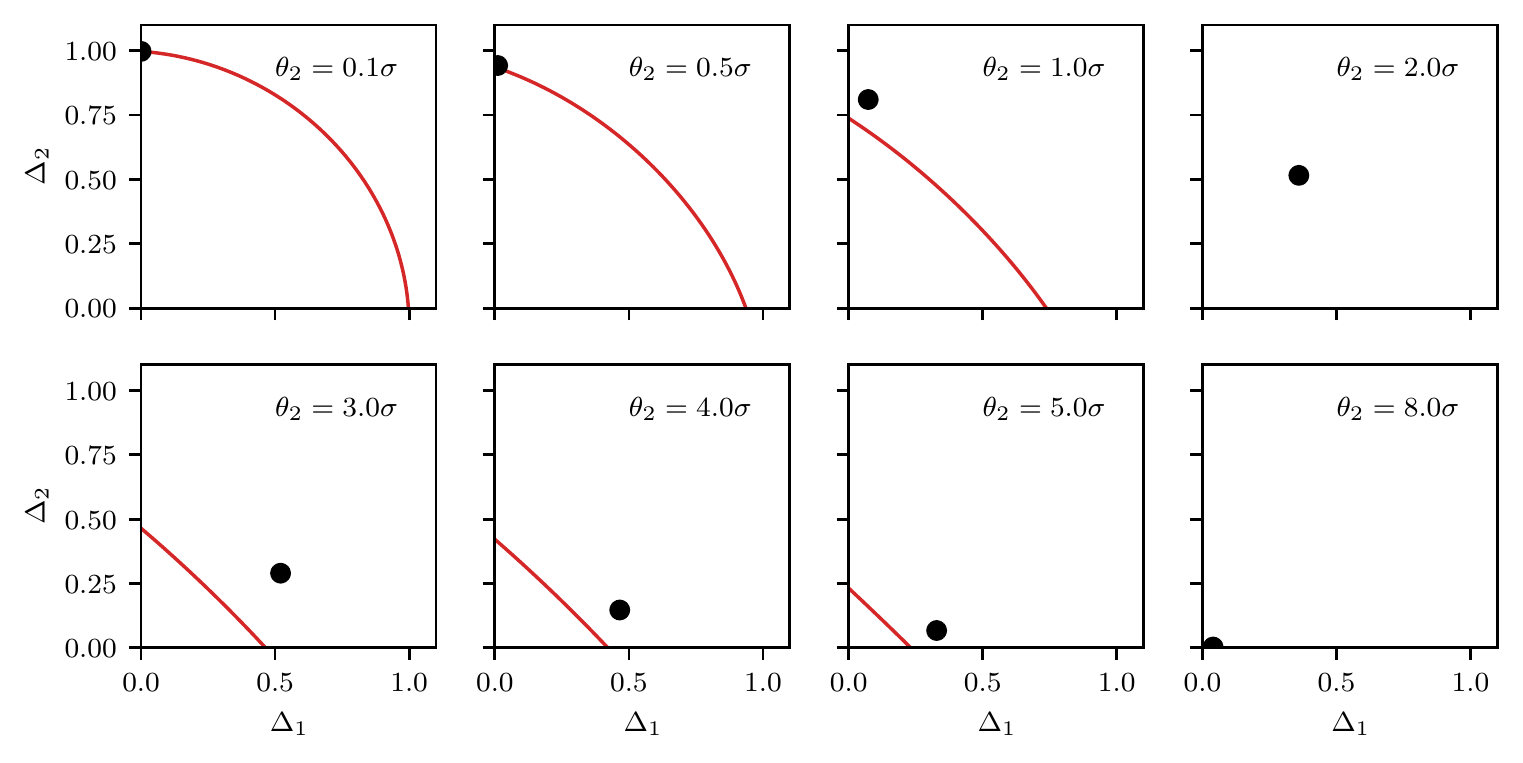}
	\caption{ 
		The IRTR and the information regrets of direct imaging with different separations.
		The black circles represent the combination \((\Delta_1, \Delta_2)\) for direct imaging. 
		The regions below the IRTR (the red solid curve) are unattainable by quantum measurements.
		For \(\theta_2=2\sigma\) (the 4th panel), \(\tilde c\) equals to zero and thus there is no IRTR.
		For \(\theta_2=8\sigma\) (the 8th panel), \(\tilde c\) is so small that the IRTR is invisible.
	}
	\label{fig:IRTR_direct_imaging}
\end{figure*}

Using the IRTR, we can obtain an intuitive picture of the possible improvements of the quantum measurement for the joint estimation of the centroid and the separation of two incoherent point sources.
It can be seen from Fig.~\ref{fig:IRTR_direct_imaging} that, for small separations (e.g., see the first three panels in Fig.~\ref{fig:IRTR_direct_imaging}), although direct imaging has a poor performance for estimating the separation, it already attains or get close to the IRTR.
Therefore, in such cases, any substantial decrease of the information regret for estimating the separation must be obtained at the cost of increasing the information regret for the centroid, in comparison with direct imaging.
When the separation of two point sources is close or equal to \(2\sigma\) (see the \(4\)th panel in Fig.~\ref{fig:IRTR_direct_imaging}), the incompatibility coefficient \(\tilde c\) is very small, indicating that it is possible to get a better joint estimation of the centroid and the separation than direct imaging.
When the separation is large enough (see the \(8\)th panel in Fig.~\ref{fig:IRTR_direct_imaging}), the information regrets for both the centroid and the separation are small.

\subsection{SPADE}
We now consider the SPADE measurement introduced in Ref.~\cite{Tsang2016b}, which takes the Hermite-Gaussian modes as the basis to decompose and measure the image-plane optical field.
We set the origin of the coordinate system on the image plane to the origin of the Hermite-Gaussian modes.
The wave function of the \(q\)-th Hermite-Gaussian mode with the same characteristic length \(\sigma\) as the Gaussian point-spread function is give by
\begin{align} \label{eq:hermite_gaussian_mode}
	\phi_q(x) = \qty(\frac{1}{2 \pi \sigma^2})^{1/4}
	\frac1{\sqrt{2^q q!}}
	H_q\qty(\frac{x}{\sqrt2 \sigma})
	\exp(-\frac{x^2}{4\sigma^2}),
\end{align} 
where \(H_q\) is the Hermite polynomial.
For the Gaussian point-spread function, the probability of obtaining the outcome \(q\) corresponding to the \(q\)-th Hermite-Gaussian mode state \(\ket{\phi_q}\) is given by 
\begin{align} 
	p(q) &= \frac12 |\ip{\phi_q}{\psi_1}|^2 + \frac12 |\ip{\phi_q}{\psi_2}|^2 \nonumber \\
	&= \frac{1}{2q!} \qty[\alpha_1^{2q} \exp(- \alpha_1^2) + \alpha_2^{2q} \exp(- \alpha_2^2)] \label{eq:hg_probability}
\end{align}
with \(\alpha_1 = X_1 / 2 \sigma = (\theta_1 - \theta_2 / 2) / 2 \sigma\) and \(\alpha_2 = X_2 / 2 \sigma = (\theta_1 + \theta_2 / 2) / 2 \sigma\).
In Ref.~\cite{Tsang2016b}, assuming that the SPADE device is aligned with the centroid of two incoherent point sources (i.e., \(\theta_1 = 0\)) such that \(\alpha_1 = - \alpha_2\) in Eq.~\eqref{eq:hg_probability}, it was shown that the SPADE is optimal for estimating the separation of the two point sources, in the sense that the information regret for \(\theta_2\) vanishes.
This optimal strategy requires the prior information or preliminary estimation of the centroid of two incoherent point sources.

\begin{figure}[tb]
	\centering
	\includegraphics[]{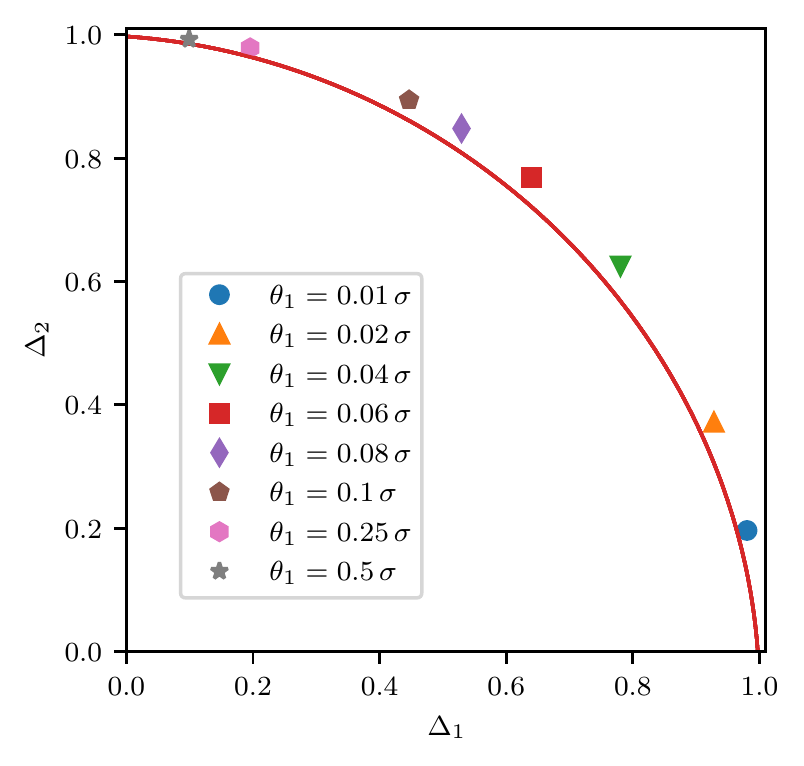}
	\caption{
		The IRTR (the red solid curve) and the information regrets of the SPADE measurement for estimating the centroid and separation of two incoherent point sources.
		Here, the information regrets are numerically computed at \(\theta_2 = 0.1 \sigma\).
	}
	\label{fig:spade}
\end{figure}

In fact, the SPADE also extracts Fisher information about the centroid from the one-photon state when there is a misalignment between the origin of the SPADE device and the centroid of the two incoherent point sources.
In Fig.~\ref{fig:spade}, we numerically compute and plot the information regrets of the SPADE for estimating \(\theta_1\) and \(\theta_2\) as well as the IRTR, for a small separation \(\theta_2=0.1\sigma\) and different values for the centroid \(\theta_1\).
We can see that, for this small separation, the information regret for \(\theta_2\) increases but the information regret for \(\theta_1\) decreases as the misalignment increases.
This means that, for two closely-placed incoherent point sources, the SPADE has a good performance for estimating the separation, when it is aligned with the centroid of the two sources, and has a good performance for estimating the centroid, when the misalignment becomes large.
The latter can be understood by considering the limit case that the origin of the Hermite-Gaussian modes is distant from the centroid of the two close point sources, i.e., \(\theta_1 \gg \theta_2\).
In such case, the probability of the SPADE measurement outcome is approximately given by 
\begin{equation}
	p(q) \approx \frac1{q!} \qty(\frac{\theta_1}{2\sigma})^{2q} \exp(- \frac{\theta_1^2}{4\sigma^2}),
\end{equation}  
meaning that the effect of the SPADE for two close incoherent point sources is like that for one point sources with twice intensity.
The Fisher information of the above probability distribution about the parameter \(\theta_1\) is \(\sigma^{-2}\), which equals its quantum limit.
Therefore, the SPADE can also be used to estimate the centroid for two closely-placed incoherent point sources.

\subsection{Random POVMs on the minimal subspace}
Note that the Hilbert space associated with the one-photon state given by Eq.~\eqref{eq:one_photon} is infinite-dimensional. 
Both the (continuum) direct imaging and the (complete) SPADE have an infinite number of possible outcomes.
Nevertheless, the FIM as well as the CRB is only relevant to the minimal subspace \(\mathcal S_\theta\) supporting the density operators and their derivatives;
This subspace \(\mathcal S_\theta\) is spanned by the state vectors \(\ket{\psi_j}\) and their derivatives \(\pdv*{\ket{\psi_j}}{X_j}\) for \(j=1,2\).
According to Naimark's dilation theorem~\cite[Section~9-6]{Peres1995}, the optimization over all quantum measurements on the underlying Hilbert space can be equivalently done by optimizing over the POVMs on the four-dimensional subspace \(\mathcal S_\theta\).

The subscript of \(\mathcal S_\theta\) indicates that this subspace depends on the value of the vector parameter \(\theta=(\theta_1, \theta_2)\), which is to be estimated.
The POVMs on \(\mathcal S_\theta\) also in general depends on \(\theta\).
This fact does not make the study of the POVMs on \(\mathcal S_\theta\) useless, due to the following reasons. 
First, remind that the (quantum) Fisher information is defined at a fiducial parameter point and measures the distinguishability about the parameters for the quantum states in the first order infinitesimal neighborhood of the fiducial state~\cite{Braunstein1994,Braunstein1996}.
The distinguishability characterized by the FIM under a \(\theta\)-dependent POVM could be attained with the prior information about the fiducial parameter point, which may be obtained by a ``pre-estimation'' stage.
Second, it is possible to implement a \(\theta\)-dependent POVM on \(\mathcal S_\theta\) through a \(\theta\)-independent POVM on the whole Hilbert space, although a systematic method to such a task is still lacking.

We now investigate the performance of randomly-generated POVMs on \(\mathcal S_\theta\) in extracting the Fisher information for both the centroid and the separation of two incoherent point sources.
For the optimization of Fisher information, it is sufficient to consider rank-1 POVMs, because all POVMs can be obtained by coarse graining of rank-1 POVMs and coarse graining cannot increase the Fisher information.
For simplicity, we only consider the orthogonal projection measurement on \(\mathcal S_\theta\) in this work and leave the optimization over general rank-1 POVMs in future work.
Furthermore, since the point-spread function is assumed to be real-valued in this work, we only consider the spatial mode whose wave functions are real.

The random orthogonal projection measurement on \(\mathcal S_\theta\) are produced as follows.
First, we apply the Gram-Schmidt process on the set \(\{\psi_1(x), \psi_2(x), \pdv{\psi_1(x)}{X_1}, \pdv{\psi_2(x)}{X_2} \}\) to get an orthonormal basis \(\qty{\phi_j(x) \mid j=1,2,3,4}\) for \(\mathcal S_\theta\).
Note that \(\phi_j(x)\) are all real-valued, as the wave functions \(\psi_1(x)\) and \(\psi_2(x)\) are real-valued.
Then, we generate ten thousand random four-dimensional orthogonal matrices~\footnote{We utilize the function ``scipy.stats.ortho\_group'' from the open-source Python library ``SciPy'' to generator the random orthogonal matrices.} with the Haar measure~\cite{Mezzadri2007}. 

\begin{figure}[tb]
	\centering
	\includegraphics[]{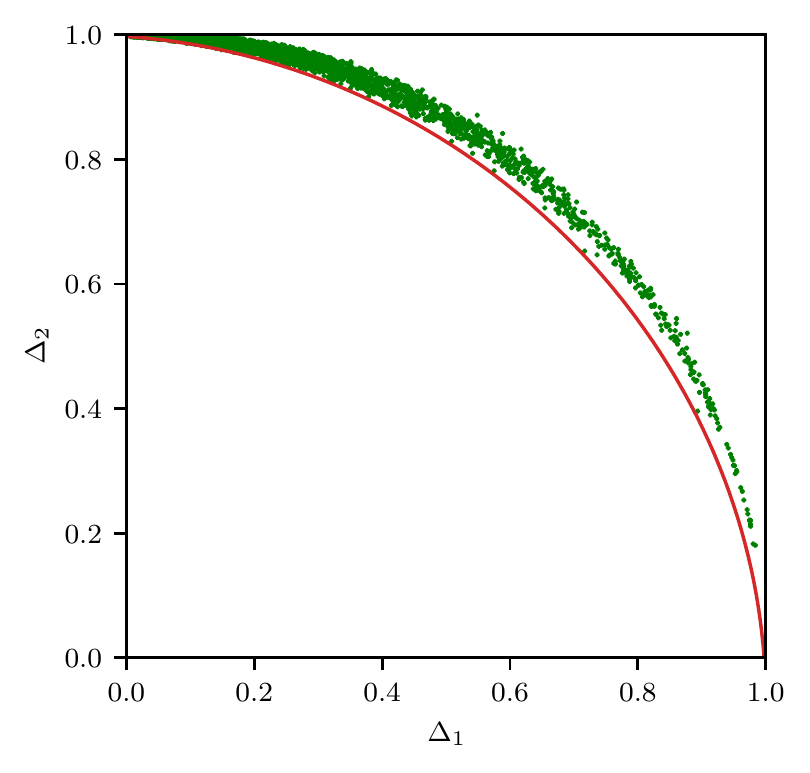}
	\caption{\label{fig:random}
	The IRTR (the red solid curve) and the information regrets of random orthogonal projection measurements on the relevant subspace.
	We generated ten thousand random real-valued orthonormal bases with the Haar measure.
	Here, \(\theta_2 = 0.1\sigma\).
	}
\end{figure}

Figure~\ref{fig:random} plots the NRS information regrets as well as the IRTR for the ten thousand random orthogonal projection measurement mentioned above, where the separation of two incoherent point sources are chosen to be \(\theta_2 = 0.1 \sigma\).
From the distributions of the information regrets \(\Delta_1\) and \(\Delta_2\), we can see that the  orthogonal projection measurement with real-valued-function on the subspace \(\mathcal S_\theta\) nearly attain the IRTR.

\section{Conclusion} \label{sec:conclusion}
In this work, we have used the IRTR to investigate the estimation of the centroid and the separation of two incoherent weak optical point sources.
We have obtained the analytic expression of the incompatibility coefficients, with which we can have an intuitive picture of the possible improvement of a measurement for estimating the centroid and the separation.
When the incompatibility coefficient is close or equal to its maximum value, corresponding to very small separations, we demonstrated that the optimal measurement for the separation must extract little information about the centroid, and vice versa.


\begin{acknowledgments}
This work is supported by the National Natural Science Foundation of China (Grants No. 61871162, No. 11805048, and No. 11935012).
\end{acknowledgments}

\appendix
\section{Calculation of the incompatibility measure} 
\label{sec:calculation_incomp}
Here, we give the detailed derivations of the expression Eq.~\eqref{eq:incompatibility} and Eq.~\eqref{eq:incompatibility_gaussian} for the incompatibility coefficient \(\tilde c\).
According to the definition of \(\tilde c\), we need the SLD operators \(L_1\) and \(L_2\) for \(\theta_1\) and \(\theta_2\).
Following Ref.~\cite{Tsang2016b}, the SLD operators can be represented as 4-dimensional matrices with an orthonormal basis for the subspace that supporting the density operators and its derivatives with respect to \(\theta_1\) and \(\theta_2\).
A specific orthonormal basis is given by~\cite{Tsang2016b}
\begin{align}
	\ket{e_1} & = \frac{1}{\sqrt{2 (1 - \delta)}} \qty(\ket{\psi_1} - \ket{\psi_2}),\\
	\ket{e_2} & = \frac{1}{\sqrt{2 (1 + \delta)}} \qty(\ket{\psi_1} + \ket{\psi_2}),\\
	\ket{e_3} & = \frac{1}{\eta_3}\qty[
  		\frac{1}{\sqrt2}\qty(\pdv{\ket{\psi_1}}{X_1} + \pdv{\ket{\psi_2}}{X_2}) - \frac{\gamma}{\sqrt{1-\delta}} \ket{e_1}
  	],\\
  	\ket{e_4} & = \frac{1}{\eta_4}\qty[
  		\frac{1}{\sqrt2}\qty(\pdv{\ket{\psi_1}}{X_1} - \pdv{\ket{\psi_2}}{X_2}) + \frac{\gamma}{\sqrt{1+\delta}} \ket{e_2}
  	],
\end{align}
where \(\kappa\), \(\gamma\), and \(\beta\) are given by Eq.~\eqref{eq:kappa}, Eq.~\eqref{eq:gamma}, and Eq.~\eqref{eq:beta}, respectively, other coefficients are defined by 
\begin{align}
	\delta 	& := \int \dd x\,\psi(x-X_1) \psi(x-X_2), \\
	\eta_3 	& :=\sqrt{\kappa + \beta - \frac{\gamma^2}{1 - \delta}},\\
	\eta_4 	& :=\sqrt{\kappa - \beta - \frac{\gamma^2}{1 + \delta}}.
\end{align}
With this basis, the density matrix for the image-plane one-photon state is given by
\begin{equation}
	\rho = \mqty(
		\frac{1-\delta}{2} & 0 & 0 & 0 \\ 
		0 & \frac{1+\delta}{2} & 0 & 0 \\
		0 & 0 & 0 & 0 \\
		0 & 0 & 0 & 0 
	).
\end{equation}
The SLD operators with respect to the centroid \(\theta_1\) and separation \(\theta_2\) are represented by
\begin{align}
	L_1 & = \mqty(
		0 & \frac{2\gamma\delta}{\sqrt{1-\delta^{2}}} & 0 & \frac{2\eta_4}{\sqrt{1-\delta}}\\
		\frac{2\gamma\delta}{\sqrt{1-\delta^{2}}} & 0 & \frac{2\eta_3}{\sqrt{1+\delta}} & 0\\
		0 & \frac{2\eta_3}{\sqrt{1+\delta}} & 0 & 0\\
		\frac{2\eta_4}{\sqrt{1-\delta}} & 0 & 0 & 0
	), \\
	L_2 & = \mqty(
		\frac{-\gamma}{1-\delta} & 0 & \frac{-\eta_3}{\sqrt{1-\delta}} & 0\\
		0 & \frac{\gamma}{1+\delta} & 0 & \frac{-\eta_4}{\sqrt{1+\delta}}\\
		\frac{-\eta_3}{\sqrt{1-\delta}} & 0 & 0 & 0\\
		0 & \frac{-\eta_4}{\sqrt{1+\delta}} & 0 & 0
	).
\end{align}

Because the matrices for \(\rho\), \(L_1\), and \(L_2\) are all real-valued, it is easy to see that 
\begin{equation}
	\qty|\tr([L_1, L_2]\rho)| = \qty|\Im\tr(L_1 L_2\rho)| = 0,
\end{equation}
implying that the quantity \(c\) defined by Eq.~\eqref{eq:c} vanishes.
By calculating the eigenvalues of \(\sqrt\rho [L_1,L_2] \sqrt\rho\), it can be shown that 
\begin{equation}
	\tr\abs{\sqrt\rho [L_1, L_2] \sqrt\rho} = 4 \abs{\beta},
\end{equation}
which implies Eq.~\eqref{eq:incompatibility}.
For the Gaussian point-spread function given by Eq.~\eqref{eq:psf_gaussian}, it can be shown that
\begin{align}
	\kappa  = \frac{1}{4\sigma^2}, \quad
	\gamma 	= -\frac{\theta_2}{4 \sigma^2} \exp(-\frac{\theta_2^2}{8 \sigma ^2}),
\end{align}
and
\begin{align}
	\beta = -\frac{\theta_2^2-4 \sigma^2}{16 \sigma^4} \exp(-\frac{\theta_2^2}{8 \sigma^2}).
\end{align}
Substituting the above expression of \(\kappa\), \(\gamma\), and \(\beta\) into Eq.~\eqref{eq:incompatibility}, we get Eq.~\eqref{eq:incompatibility_gaussian}.

\bibliography{../../../../Workplace/wiki}
\end{document}